\documentclass[reprint,aps,pra,superscriptaddress,amsmath,amssymb
]{revtex4-2}
\usepackage[utf8]{inputenc}
\usepackage{bbm}
\usepackage{graphicx}
\usepackage{dcolumn}
\usepackage{bm}
\usepackage{mathrsfs}
\usepackage{placeins} 
\usepackage{amsfonts}
\usepackage{color}
\usepackage{xcolor}
\usepackage{braket,verbatim}
\usepackage[english]{babel}
\usepackage{lipsum} 
\addto\captionsenglish{}
\usepackage{url}
\definecolor{darkblue}{rgb}{0,0.3,0.7}
\usepackage{hyperref}
\hypersetup{
colorlinks=true,
linkcolor=darkblue,
filecolor=blue,
citecolor=darkblue,  
urlcolor=darkblue,
}

\DeclareMathOperator{\Tr}{Tr}

\newcommand{\imag}{\mathrm{i}}
\newcommand{\bt}{\boldsymbol{t}}
\newcommand{\cI}{\mathcal{I}}
\newcommand{\cL}{\mathcal{L}}
\newcommand{\cC}{\mathcal{C}}
\newcommand{\cP}{\mathcal{P}}
\newcommand{\cH}{\mathcal{H}}
\newcommand{\cD}{\bm{\mathcal{D}}}
\newcommand{\bbE}{\mathbb{E}}

\definecolor{jmlBlue}{RGB}{0,163,224}

\definecolor{lightRed}{RGB}{255,64,64}

\definecolor{ajpGreen}{RGB}{0,102,0}

\begin{document}

\title{Comparing Homodyne and Heterodyne Tomography of Quantum States of Light}

\author{Rhea P. Fernandes}
\thanks{Equal contribution}
\email{ferna416@purdue.edu}
\affiliation{Elmore Family School of Electrical and Computer Engineering and Purdue Quantum Science and Engineering Institute, Purdue University, West Lafayette, Indiana 47907, USA}

\author{Andrew J. Pizzimenti}
\thanks{Equal contribution}
\email{ajpizzimenti@arizona.edu}
\affiliation{Wyant College of Optical Sciences, The University of Arizona, Tucson, AZ 85721, USA}

\author{Christos N. Gagatsos}
\email{cgagatsos@arizona.edu}
\affiliation{Department of Electrical and Computer Engineering, The University of Arizona, Tucson, Arizona, 85721, USA}
\affiliation{Wyant College of Optical Sciences, The University of Arizona, Tucson, AZ 85721, USA}
\affiliation{Program in Applied Mathematics, The University of Arizona, Tucson, Arizona 85721, USA}

\author{Joseph M. Lukens}
\email{jlukens@purdue.edu}
\affiliation{Elmore Family School of Electrical and Computer Engineering and Purdue Quantum Science and Engineering Institute, Purdue University, West Lafayette, Indiana 47907, USA}
\affiliation{Quantum Information Science Section, Oak Ridge National Laboratory, Oak Ridge, Tennessee 37831, USA}

\date{\today}

\begin{abstract}
Non-Gaussian quantum states are critical resources in photonic quantum information processing, rendering their generation and characterization of increasing importance in quantum optics. In this work, we theoretically and numerically analyze the relative efficiency of homodyne versus heterodyne measurements for reconstructing non-Gaussian states, a major outstanding question in continuous-variable tomography. Combining a Fisher information-based formalism with simulated experiments, we find homodyne tomography to outperform heterodyne measurements for all non-Gaussian states tested, although the separation between the two modalities proves significantly narrower than suggested by the asymptotic Cram\'{e}r--Rao lower bound.
Our results should find use for optimizing measurement strategies in practical continuous-variable quantum systems.
\end{abstract}

\maketitle

\section{Introduction}\label{sec:Introduction}
Continuous-variable (CV) quantum optics presents valuable opportunities for quantum information processing (QIP)~\cite{Braunstein2005}. In contrast to discrete-variable (DV) photonic quantum information, where historically the most common qubit has been a single photon spread over two modes~\cite{Knill2001, Kok2007}, the fundamental units of information in CV approaches are ``qumodes''---single modes that can carry many photons~\cite{Barbosa2003}. CV quantum systems furnish unique capabilities complementary to more established DV formalisms. In quantum communications, CV quantum key distribution can be realized with inexpensive off-the-shelf detection systems from classical coherent communications~\cite{Grosshans2002, Laudenbach2018}, while quantum sensors  based on squeezed light enable quantum sensitivity enhancements even for extremely bright fields~\cite{Aasi2013, Lawrie2019, Yu2020}. Computationally, Gaussian boson sampling (GBS) unlocks quantum advantages more readily than DV boson sampling due to deterministic state generation~\cite{Hamilton2017, Kruse2019}, while Gottesman--Knill--Preskill (GKP) encodings provide a compelling path for error correction~\cite{Gottesman2001}.

Within this CV ecosystem, the distinction between Gaussian and non-Gaussian quantum states has proven useful both theoretically and experimentally. Defined by Wigner functions that are themselves Gaussian~\cite{Weedbrook2012}, Gaussian states can be generated deterministically with existing technology and encompass such optical workhorses as coherent, thermal, and squeezed states. Yet Gaussian states can be efficiently simulated classically~\cite{Bartlett2002}, making non-Gaussian resources (either states or measurements) critical for universal QIP~\cite{Lloyd1999, Walschaers2021}. Beyond arguably straightforward examples such as single-photon Fock states~\cite{Lvovsky2001}, non-Gaussian photonic states are notoriously difficult to generate, but experimental progress is accelerating. Indeed, foundational cat state examples based on photon subtraction~\cite{Ourjoumtsev2006, Neergaard2006, Wakui2007, Ourjoumtsev2007, Takahashi2008, Gerrits2010} have recently been extended to highly multimode non-Gaussian states~\cite{Ra2020}, and GBS experiments (for which photon detection represents the needed non-Gaussian resource)~\cite{Zhong2019, Paesani2019, Zhong2020, Arrazola2021, Zhong2021, Thekkadath2022}---including a recent demonstration of an optical GKP qubit~\cite{Larsen2025}---inspire optimism for a variety of configurations for tailored non-Gaussian state generation based on squeezed inputs, multiport interferometers, and partial mode detection~\cite{Gagatsos2019, Sabapathy2019, Quesada2019, Su2019, Tzitrin2020, Walschaers2020, Pizzimenti2021,Pizzimenti2024}.

As such non-Gaussian CV states become increasingly common, so too does the need for their efficient characterization. Homodyne and heterodyne tomography represent natural choices---the former detecting a single field quadrature per mode per state copy, the latter 
detecting two quadratures in parallel. Both approaches are tomographically complete and thus guaranteed to recover the ground truth asymptotically. However, the \emph{efficiency} with which they do so can depend strongly on the state in question. Although previous research has explored this question for Gaussian states~\cite{Rehacek2015, Muller2016, Teo2017}, to our knowledge no prior work has considered the general case of arbitrary CV states. Thus, there exists no answer to the question: \emph{For a given non-Gaussian state, is homodyne or heterodyne tomography more efficient?}

In this paper, we explore this question systematically in both theory and simulation. After deriving the classical Fisher information (CFI) associated with either homodyne or heterodyne measurements on arbitrary single-mode quantum states, we perform maximum likelihood estimation (MLE) on simulated data generated by a variety of quantum systems, including random mixed states and selected Gaussian and non-Gaussian examples up to a photon number cutoff of 10.

Interestingly, while the homodyne simulations show good agreement with expectations from the Cram\'{e}r--Rao lower bound (CRLB), estimation errors for the heterodyne tests deviate significantly from the CRLB even up to 10$^9$ state copies. Notwithstanding, for all states examined, both the CFI and empirical MLE tests convincingly highlight the effectiveness of homodyne over heterodyne measurements in CV state estimation---not just for non-Gaussian states, but also Gaussian states parametrized generically (i.e., without enforcing Gaussianity). Collectively, our results should prove useful in the continued progress of CV quantum information, furnishing important tools and insights for efficient tomographic schemes.
\section{Theory}\label{sec:theory}
\subsection{Background}
\label{sec:background}
\textit{CV state representations.---}Quantum state tomography (QST) is the process by which a quantum state is reconstructed using measurements on a collection of identical copies ~\cite{RevModPhys.81.299, PhysRevA.64.052312}. 
To uniquely identify the state, it is desirable for the measurements to be tomographically complete, meaning they are sensitive to all unknown parameters that describe the quantum system. 
In the CV regime in which the signals of interest can contain an uncertain number of photons, field quadrature measurements are often performed. The phase of this quantum measurement is set by a local oscillator (LO).

For this study, we consider ground truth quantum states in a single optical mode (spatial, temporal, and polarization), each described by a density matrix in the Fock basis of the form
\begin{equation}
\label{eq:rho}
\rho = \sum_{m=0}^\infty\sum_{n=0}^\infty \rho_{mn}\ket{m}\bra{n}.
\end{equation}
Equivalently, $\rho$ can be represented via quasiprobability distributions, such as the Wigner function~\cite{Lvovsky2009}
\begin{equation}
\label{eq:Wigner}
W(x,p)=\frac{1}{\pi}\int_{-\infty}^\infty dy\,\braket{x-y|\rho|x+y}e^{-2\imag py}
\end{equation}
and Husimi $Q$ function~\cite{Orszag2024}
\begin{equation}
\label{eq:Husimi}
Q(\alpha)=\frac{1}{\pi}\braket{\alpha|\rho|\alpha}.
\end{equation}
We adopt the $\hbar=1$ convention where $\ket{x}$ is the $\theta=0$ version of the general quadrature eigenstate $|x_\theta\rangle$ at angle $\theta$~\cite{Lvovsky2009}:
\begin{equation}
\label{eq:xFock}
\braket{n|x_\theta} = \frac{1}{\pi^{1/4}\sqrt{2^n n!}}e^{\imag n\theta}e^{-x^2/2}H_n(x),
\end{equation}
with $H_n(x)$ a Hermite polynomial. The coherent state $\ket{\alpha}$ follows the standard form
\begin{equation}
\label{eq:alphaFock}
\braket{n|\alpha} = \frac{1}{\sqrt{n!}}e^{-|\alpha|^2/2}\alpha^n,
\end{equation}
also expressed in the Fock basis.
The QST problem can therefore be summarized as obtaining an estimate of any of the above equivalent state representations [Eqs.~(\ref{eq:rho}--\ref{eq:Husimi})] through measurements on a collection of copies. 

\begin{figure}[tb!]
    \centering
    \includegraphics[width= 8cm]{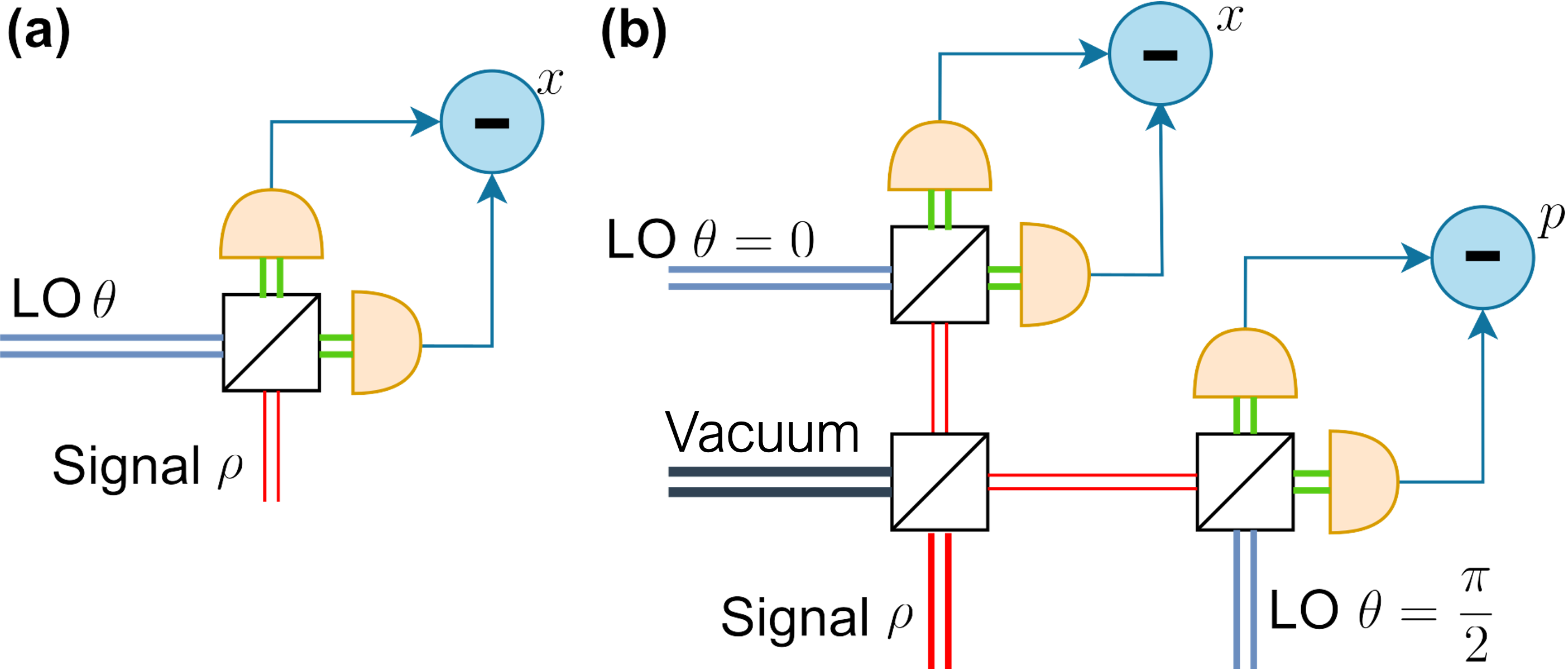}
    \caption{Schematic of (a) homodyne detection and (b) heterodyne detection.}
    \label{fig:hom}
\end{figure}

\textit{Homodyne detection.---}Homodyne detection directly measures the phase-dependent quadrature of the optical  field 
by mixing the state with an LO with phase $\theta$ on a beamsplitter ~\cite{PhysRevA.40.2847, scully1997quantum,walls2008quantum}. Balanced detection is then performed on the two outputs. In the typical case, the beam splitter has a 1:1 split ratio (50\% transmission and 50\% reflection). Figure~\ref{fig:hom}(a) shows the basic configuration for homodyne detection, where we assume perfect matching of the spatio-polarization and temporal modes for the LO and signal. For each copy of the quantum state, this returns one measurement result $x$ with a probability density function (PDF)
~\cite{RevModPhys.81.299,Chapman:22}
\begin{equation}
\begin{split}
\label{eq:homPDF}
f_\theta(x) & = \sum_{m=0}^\infty \sum_{n=0}^\infty \rho_{mn} \frac{e^{-\imag (m-n)\theta}}{\sqrt{\pi m! n! 2^{m+n}}} e^{-x^2}H_m(x)H_n(x) \\
& = \braket{x_\theta|\rho|x_\theta}.
\end{split}
\end{equation}
For $\rho=\ket{0}\bra{0}$, Eq.~(\ref{eq:homPDF}) reduces to a normal distribution with variance $\braket{\Delta x^2}=\frac{1}{2}$ as required under the $\hbar=1$ convention.

Significantly, the homodyne PDF is directly related to the Wigner function according to
\begin{equation}
\label{eq:HOMwigner}
f_\theta(x)=\int_{-\infty}^\infty dp\,W(x\cos\theta-p\sin\theta,x\sin\theta+p\cos\theta),
\end{equation}
so that homodyne tomography can be viewed as sampling the Wigner function at angles $\theta$ determined by the LO~\cite{Lvovsky2009}.

\textit{Heterodyne detection.---}Heterodyne detection~\cite{Walker1986MultiportHD, Chapman:22} first splits the signal in two with a 50/50 beamsplitter and then performs homodyne detection separately on the two outputs, one with an LO phase $\theta=0$ and the other with phase $\theta=\frac{\pi}{2}$. 
Figure~\ref{fig:hom}(b) shows the basic configuration of heterodyne detection. In this case, the PDF for the ordered pair $(x,p)$ is~\cite{Richter1998,Chapman:22}
\begin{equation}
\begin{split}
\label{eq:HETpdf}
g(x,p) & =\sum_{m=0}^\infty \sum_{n=0}^\infty\rho_{mn}  \frac{e^{-(x^2+p^2 )}}{\pi \sqrt{m!n!}}   (x-\imag p)^m (x+\imag p)^n \\
& = \frac{1}{\pi} \braket{x+\imag p|\rho|x+\imag p},
\end{split}
\end{equation}
which is precisely the Husimi $Q$ function [Eq.~(\ref{eq:Husimi})] with coherent state parameter $\alpha=x+\imag p$~\cite{Leonhardt1993,Richter1998}.

\textit{Previous work.---}In light of each modality's direct mathematical relationship to a complete quasiprobability representation ($W$ or $Q$), both homodyne and heterodyne tomography are tomographically complete and therefore sufficient for estimating arbitrary ground truth quantum states. However, the efficiency at which they do so---quantified as error versus the number of state copies---need not be the same. Due to the additional vacuum noise injected by the first beasmplitter in heterodyne measurements [black lines in Fig.~\ref{fig:hom}(b)], 
it is our experience that many researchers in quantum optics seem to prefer homodyne tomography for its perceived lower noise. Interestingly, though, a handful of theoretical and experimental studies have actually found heterodyne measurements to outperform homodyne, reaching lower errors in estimating the covariance matrix for most---though not all---Gaussian states~\cite{Rehacek2015, Muller2016, Teo2017}. 

Such findings reveal the naivety of a comprehensive preference for homodyne measurements, yet the results themselves remain fairly narrow. Not only do they apply to Gaussian ground truth states alone, but they also assume the experimenter \emph{knows} the states are Gaussian and can model them as such. This latter condition is significant, as it precludes the applicability of the results in \cite{Rehacek2015, Muller2016, Teo2017} to the typical QST paradigm 
where the parametrized estimator $\hat{\rho}$ is  
sufficiently general to express all possible quantum states; i.e., even when the ground truth is \emph{in fact} Gaussian, QST is asked to confirm Gaussianity, not assume it. 







\subsection{Approach}
\label{sec:approach}
To benchmark the performance of homodyne and heterodyne detection, we utilize a method based on the CFI. Suppose our system of interest can be described by a vector of parameters $\bt$, on which measurements are performed to produce a dataset $\cD$ according to a probability distribution $\Pr(\cD|\bt)$.
The CFI $\cI$ quantifies the amount of information $\Pr(\cD|\bt)$ gives about $\bt$, while the inverse lower bounds the variance of any unbiased estimator $\hat{\bt}$ through  the CRLB. Throughout this work, we reserve  caret notation for any estimator and plain symbols for the ground truth, irrespective of the quantity's status as an operator; e.g., $\hat{\bt}$ is the estimate for $\bt$, $\hat{\rho}$ for $\rho$. 

The likelihood is defined as $\cL(\bt)\propto\Pr(\cD|\bt)$, i.e., equal to the conditional probability distribution up to irrelevant proportionality factors that depend on $\cD$ only. 
Then the CFI matrix with elements 
\begin{equation}
    \cI_{ij} = \bbE\left[ \left(\frac{\partial}{\partial t_i} \ln\cL(\bt) \right)\left(\frac{\partial}{\partial t_j} \ln\cL(\bt)\right)\right]
\end{equation}
lower-bounds the covariance matrix of  any unbiased estimator $\hat{\bt}$ via the CRLB:
\begin{align}
\label{eq:CRLB}
    \text{cov}\left(\hat{\bt} \right) \succeq\mathcal{I}^{-1}
\end{align}
in the positive semdefinite sense. Here and in all that follows, expectations are defined assuming a fixed ground truth $\rho(\bt)$ so that $\bbE[\cdot] = \bbE[\cdot|\bt]$ is implied.

Truncating the formally infinite-dimensional quantum state in Eq.~(\ref{eq:rho}) to a photon cutoff $n_c$---required to obtain a finite number of unknown parameters and consonant with ``finite-energy'' assumptions---we adopt a parametrization 
based on
the generalized Gell--Mann (GGM) matrices $\boldsymbol{\Omega}$ \cite{KIMURA2003339, Bertlmann2008}:
\begin{equation}
\label{eq:rhoGell}
\rho(\bt) = \frac{\mathbbm{1}}{d} + \bt \cdot \boldsymbol{\Omega} = \frac{\mathbbm{1}}{d} + \sum_{i=1}^{d^2-1} t_i \Omega_i,
\end{equation}
where $d=n_c+1$ is the dimension of the Hilbert space $\cH_d$ of interest (including vacuum), and the $d^2-1$ GGM matrices $\boldsymbol{\Omega} = (\Omega_1,...,\Omega_{d^2-1})$ are defined 
according to three classes:
\begin{enumerate}
\item For $i\in\set{1,...,\frac{1}{2}d(d-1)}$, the traceless symmetric matrices for all unique $l,m$ pairings subject to $0 \leq l < m \leq d-1$:
\begin{equation} 
    \Omega_i =  |l\rangle\langle m| + |m\rangle\langle l|.
\end{equation}
\item For $i\in\set{\frac{1}{2}d(d-1)+1,...,d(d-1)}$, all unique traceless antisymmetric matrices for $0 \leq l < m \leq d-1$:
\begin{equation}
    \Omega_i = -\imag |l\rangle\langle m| +\imag |m\rangle\langle l|.
\end{equation}
\item For $i\in\set{d(d-1)+1,...,d^2-1}$, the $d-1$ diagonal matrices for $1 \leq l\leq d-1$:
\begin{equation}
    \Omega_i =  \sqrt{\frac{2}{l(l+1)}} \left(\sum_{m=1}^l |m\rangle\langle m| - l |l\rangle\langle l| \right).
\end{equation}
\end{enumerate}
Under these definitions, the orthogonality condition $\Tr \Omega_i\Omega_j = 2\delta_{ij}$ holds, which in turn allows the CRLB to be re-expressed in terms of
the squared Frobenius error between the estimator $\hat{\rho}$ and ground truth $\rho$. Observing that
\begin{equation}
\label{eq:Fro}
\|\hat{\rho}-\rho\|_F^2 = \Tr (\hat{\rho}-\rho)^\dagger (\hat{\rho}-\rho) = 2\sum_{i=1}^{d^2-1}(\hat{t}_i-t_i)^2,
\end{equation}
and taking the expectation, yields
\begin{equation}
\label{eq:ExpFro}
\bbE[\|\hat{\rho}-\rho\|_F^2] = 2\sum_{i=1}^{d^2-1} \bbE[(\hat{t}_i-t_i)^2] = 2\Tr \text{cov}\left(\hat{\bt} \right),
\end{equation}
under the assumption that $\bbE[\hat{\bt}]=\bt$ (i.e., $\hat{\bt}$ is unbiased). The last expression can immediately be identified with the CRLB in Eq.~(\ref{eq:CRLB}), leaving the relation
\begin{equation}
\label{eq:FroCRLB}
\bbE[\|\hat{\rho}-\rho\|_F^2] \geq 2\Tr \cI^{-1},
\end{equation}
which matches previous findings of Acharya \emph{et al.}~\cite{Acharya2016, Acharya2019} with the weight matrix $G(\bt)=2\mathbbm{1}_{d^2-1}$. 

Equation~(\ref{eq:FroCRLB}) represents a particularly convenient expression for tomographic analyses. By simplifying the full matrix form of the CRLB [Eq.~(\ref{eq:CRLB})] to a scalar condition, agreement with the CRLB can be explored with a simple error measure defined on the complete quantum state. Although the specific parametrization in Eq.~(\ref{eq:rhoGell}) plays a critical role in the definition and calculation of $\cI$, the Frobenius error $\|\hat{\rho}-\rho\|_F^2$ is parametrization-independent, suggesting that the estimator $\hat{\rho}$ can be obtained with no reference to the GGM expansion coefficients $\hat{\bt}$---allowing us in Sec.~\ref{sec:simulation} to employ an efficient iterative MLE algorithm~\cite{Lvovsky2004} that computes $\hat{\rho}$ directly without an explicit basis expansion.

\subsection{Computing the CFI}
\label{sec:computeCFI}
For the PDFs and generic states $\rho$ under consideration here, we are not aware of any approach to obtain $\cI$ from direct integration. Consequently, we adopt a numerical workflow introduced in \cite{Rehacek2015}. Considering homodyne measurements first, we suppose that $S$ LO settings $\set{\theta_1,...,\theta_S}$ are measured $K/S$ times each for a total number of state copies $K$. By discretizing the quadrature space into $N$ bins  such that $x_i=x_1+(i-1)\Delta x$ ($i\in\set{1,...,N}$), we can define a multinomial distribution for each LO phase $s\in\set{1,...,S}$ with probabilities $f_{\theta_s}(x_i)\Delta x$ [Eq.~(\ref{eq:homPDF})]. With the dataset $\cD=(n_{11},...,n_{SN})$ collecting the outcomes registered for each LO phase $s$ and bin $i$, the discretized log-likelihood becomes
\begin{equation}
\begin{split}
\ln\cL_\mathrm{hom} & = \sum_{s=1}^S\sum_{i=1}^N n_{si} \ln \braket{(x_i)_{\theta_s}|\rho|(x_i)_{\theta_s}},\\
& = \sum_{s=1}^S\sum_{i=1}^N n_{si} \ln \left(\frac{\mathbbm{1}}{d} + \sum_{j=1}^{d^2-1} t_j\braket{(x_i)_{\theta_s}|\Omega_j|(x_i)_{\theta_s}} \right),\label{eq:logLLhom}
\end{split}
\end{equation}
where $\sum_{i=1}^N n_{si} = K/S$.

Following the logic of \cite{Rehacek2015}, we can simplify the CFI into the form
\begin{equation}
\label{eq:homCFImatrix}
\cI_\mathrm{hom} = K \left( \frac{1}{S}\sum_{s=1}^S \cC_s^T \cP_s^{-1} \cC_s \right),
\end{equation}
with $\cC_s$ an $N\times (d^2-1)$ matrix defined by elements
\begin{equation}
\label{eq:homC}
(\cC_s)_{ij} = \braket{(x_i)_{\theta_s}|\Omega_j|(x_i)_{\theta_s}}\\
\end{equation}
and $\cP_s$ an $N\times N$ diagonal matrix with entries 
\begin{equation}
\label{eq:homP}
(\cP_s)_{ii} = \frac{1}{d} + \sum_{j=1}^{d^2-1} t_j\braket{(x_i)_{\theta_s}|\Omega_j|(x_i)_{\theta_s}}.
\end{equation}
By Eq.~(\ref{eq:homCFImatrix}), the CFI 
is therefore proportional to the average of CFIs associated with each LO phase $\theta_s$, 
with a proportionality factor equal to the total  number of state copies $K$.

The heterodyne CFI calculation proceeds similarly, but now both $x$ and $p$ are discretized, which for convenience we implement through a single index $i\in\set{1,...,N^2}$, i.e., $x_i = x_1 + \lfloor i/N\rfloor\Delta x$ and $p_i = p_1+[(i\mod N) -1)]\Delta p$. Since only one LO phase is required for heterodyne tomography, the log-likelihood is now
\begin{equation}
\begin{split}
\ln\cL_\mathrm{het} & = \sum_{i=1}^{N^2} n_i \ln \braket{x_i+\imag p_i|\rho|x_i+\imag p_i},\\
& = \sum_{i=1}^{N^2} n_i \ln \left(\frac{\mathbbm{1}}{d} + \sum_{j=1}^{d^2-1} t_j\braket{x_i+\imag p_i|\Omega_j|x_i+\imag p_i} \right),\label{eq:logLLhet}
\end{split}
\end{equation}
where $\sum_{i=1}^{N^2}n_i=K$. The matrix version of the CFI emerges as
\begin{equation}
\label{eq:hetCFImatrix}
\cI_\mathrm{het} = K \cC^T \cP^{-1} \cC,
\end{equation}
with $\cC$ an $N^2\times (d^2-1)$ matrix
\begin{equation}
\label{eq:hetC}
\cC_{ij} = \braket{x_i+\imag p_i|\Omega_j|x_i+\imag p_i}
\end{equation}
and $\cP$ an $N^2\times N^2$ diagonal matrix
\begin{equation}
\label{eq:hetP}
\cP_{ii} = \frac{1}{d} + \sum_{j=1}^{d^2-1} t_j\braket{x_i+\imag p_i|\Omega_j|x_i+\imag p_i}.
\end{equation}

To summarize, then, the procedure for calculating the CFI consists of (i)~computing the Bloch vector $\bt$ associated with the state of interest $\rho$; (ii)~defining the discretized phase space $x_i$ for homodyne or $(x_i,p_i)$ for heterodyne; (iii)~calculating the matrices $\cC$, $\cP$, and  $\cC^T\cP^{-1}\cC$; and (iv)~in the homodyne case, repeating step (iii) for all LO phases. In the limits $\Delta x, \Delta p \rightarrow 0$ and $N,S\rightarrow \infty$, the CFI will converge to the continuous definition; in practice, we ensure convergence by taking the following steps for both homodyne and heterodyne measurements and all states considered: 
\begin{enumerate}
    \item Construct the array $\mathcal A = \Tr\cI^{-1}\left(S,x_{1},\Delta x\right)$, for $S \in \{2,3,4,5,6\}\times10^2$, $x_1 \in \{-2.5, -3.75, -5, -6.5, -7.5\}$, $\Delta x \in \{1.5, 1,0.5,0.1,0.05\}$, and $N$ such that $x_N = -x_1$. For all heterodyne tests, the $p$-quadrature vector is identical to $x$: $(p_1, \Delta p, p_N) = (x_1, \Delta x, x_N)$.
    \item Calculate the percent error between $\mathcal A_{i,j,k}$ and its seven nearest neighbors in the array along the directions of increasing $S$, decreasing $x_1$, and decreasing $\Delta x$. 
    \item Find $(\tilde{\imath},\tilde{\jmath},\tilde{k})$ such that the percent error between $\mathcal A_{\tilde{\imath},\tilde{\jmath},\tilde{k}}$ and the seven nearest neighbors considered in Step $2$ is $<2\%$ simultaneously.
\end{enumerate}
The continuous limit of $\Tr \cI^{-1}$ is then well approximated by $\Tr \cI^{-1}(S_{\tilde{\imath}},x_{1_{\tilde{\jmath}}},\Delta x_{\tilde{k}})^{-1}$, since increasing the phase space boundaries, resolution, number of LO phases, or any combination thereof results in a negligible difference. We find $(S,x_1,\Delta x) = (500, -5, 0.1)$ 
sufficient in all cases.   

To the extent that the CRLB is both obeyed and tight in the regime of interest, the formalism developed in this section is in principle sufficient to answer our original question: are homodyne or heterodyne measurements more efficient to characterize a given ground truth state $\rho$? According to Eq.~(\ref{eq:FroCRLB}), one only needs to calculate $\cI_\mathrm{hom}$ and $\cI_\mathrm{het}$ following the procedures outlined here to bound the mean squared Frobenius error $\bbE[\|\hat{\rho}-\rho\|_F^2]$. However, care must be taken in drawing conclusions from Eq.~(\ref{eq:FroCRLB}), for it---like the CRLB itself---relies on \emph{asymptotic normality} \cite{Guta2011}. As discussed in previous QST contexts~\cite{Acharya2019, Scholten_2018}, both the presence of boundaries in the physical Hilbert space ($\rho\succeq 0$ and $\Tr\rho=1$) and the finite number of measurements in practice can lead to significant deviations from CRLB predictions. Consequently, for our purposes Eq.~(\ref{eq:FroCRLB}) represents a valuable starting point, with the numerical simulations explored in Sec.~\ref{sec:simulation} designed to test realistic scenarios for agreement with the ideal asymptotic regime.

\section{Simulation}\label{sec:simulation}
\subsection{Overview}
In order to compare homodyne and heterodyne measurements numerically, we simulate experiments with
each quadrature divided into $N$ bins of width $\Delta x$ or $\Delta p$, leaving discrete probabilities: $\Pr_s[i]=f_{\theta_s}(x_i)\Delta x$ ($i\in\set{1,...,N}$) for each LO phase $s$ measured with homodyne, $\Pr[i]=g(x_i,p_i)\Delta x\Delta p$ ($i\in\set{1,...,N^2}$) for heterodyne. A multinomial simulation returns vectors of counts $\cD=(n_{11},...,n_{SN})$ (homodyne) and $\cD=(n_1,...,n_{N^2})$ (heterodyne) for a total number of state copies $K_{\max}$. With these vectors in hand, we consider a specific number of copies $K\in\set{1,...,K_{\max}}$ and find the state $\hat{\rho}$ that maximizes the respective likelihood [Eq.~(\ref{eq:logLLhom}) or (\ref{eq:logLLhet})] using an iterative approach~\cite{Lvovsky2004}.
Performing each experiment $E$ times allows us to approximate the mean-squared Frobenius error as 
\begin{equation}
\label{eq:MSE}
\bbE\left[\|\hat{\rho}-\rho\|_F^2\right]\approx \frac{1}{E}\sum_{e=1}^E \|\hat{\rho}_e-\rho\|_F^2,
\end{equation}
which we then compare against the CRLB in Eq.~(\ref{eq:FroCRLB}).

\begin{table}[]
    \centering
    \begin{tabular}{|c|c|c|c|c|c|c|c|c|}
         \hline
          & $x_1$ & $\Delta x$ & $p_1$ & $\Delta p$ & $N$ & $S$ & $K_{\max}$ & $E$ \\\hline
         Homodyne & $-10$ & 0.1005 & --  & -- & 200 & 100 & $10^9$ & 10 \\
         Heterodyne & $-10$ & 0.1005 & $-10$  & 0.1005 & 200 &  -- & $10^9$ & 10 \\\hline
    \end{tabular}
    \caption{Parameters selected for numerical simulations. Initial $x$-quadrature  point $x_1$ and spacing $\Delta x$, $p$-quadrature starting point $p_1$ and spacing $\Delta p$, number of bins per quadrature $N$, number of LO phases $S$ (evenly spaced between 0 and $2\pi$), maximum number of state copies measured $K_{\max}$, and total number of experiments $E$ per quantum state.}
    \label{tab:params}
\end{table}

\begin{figure*}[bt!]
\centering
\includegraphics[width=7in]{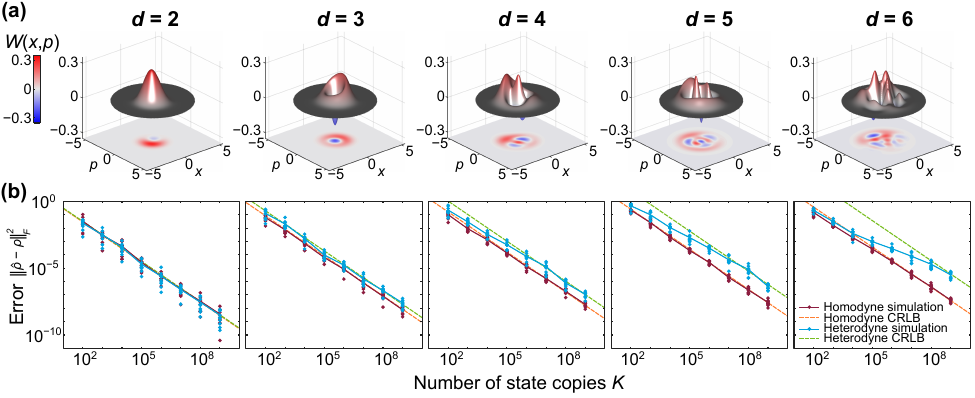}
\caption{Simulation results for random non-Gaussian states with Hilbert space dimension $d\in\{2,3,4,5,6\}$. (a) Ground truth Wigner functions and (b) corresponding estimation errors. Circles denote the Frobenius error for a specific trial, solid lines trace the mean Frobenius error of ten trials, and dashed lines give the CRLB.}
\label{fig:rand1}
\end{figure*}

Table~\ref{tab:params} lists the parameter values employed in all trials. 
Simulated experiments and MLE algorithms are run in MATLAB on a desktop computer with 256 GB of RAM and 56 dual-core processors (up to 112 threads).

\subsection{Random States}
\label{sec:random}
We first apply the developed analysis to randomly generated ground truth states possessing a set photon number cutoff $n_c$ and with purity in the range $\Tr\rho^2\in(0.70,0.95)$. 
Specifically, for each Hilbert space dimension $d\in\set{2,...,11}$ (photon cutoff $n_c\in\set{1,...,10}$), we generate one random ground truth state $\rho$ and conduct $E=10$ experiments each for homodyne and heterodyne measurements. Figure~\ref{fig:rand1} shows the results for $d\leq 6$ and Fig.~\ref{fig:rand2} for $d\geq 7$. For all dimensions $d$, the ground truth Wigner functions [Figs.~\ref{fig:rand1}(a) and \ref{fig:rand2}(a)] are highly non-Gaussian and include negative regions in the phase space.

According to the CFI, homodyne measurements are more efficient than heterodyne for all ten tested states, with the separation widening significantly as $d$ increases. Overall, the MLE results confirm these trends, yet the degree of agreement with the CRLB varies; whereas the homodyne error converges rapidly to the CRLB for all $d$, the heterodyne Frobenius error does so relatively slowly, with the $d\geq 7$ cases not yet touching the CRLB after the explored $K_{\max}=10^9$ measurements. Since the CRLB is asymptotic, such findings do not contradict the theory underpinning Eq.~(\ref{eq:FroCRLB}), but they do reaffirm the care which must be taken when applying CRLB predictions to the finite-measurement regime. 


\subsection{Tailored States}
\label{sec:tailored}
For additional comparisons between homodyne and heterodyne tomography, we next apply our formalism to user-defined ground truth states drawn from standard families of Gaussian and non-Gaussian states. For all tests, we take $d=n_c+1=11$ but otherwise consider the same parameter settings as in Table~\ref{tab:params}. We select six states for analysis:
\begin{enumerate}
\item Thermal state with $\lambda= 0.5$:
\begin{equation}
\rho = (1 - |\lambda|^2) \sum_{n=0}^{n_c} |\lambda|^{2n} \ket{n}\bra{n}.
\end{equation}
\item Coherent state $\rho=\ket{\alpha}\bra{\alpha}$ with $\alpha = 1.8$ [cf. Eq.~(\ref{eq:alphaFock})].
\item Squeezed vacuum $\rho=\ket{r}\bra{r}$ with $r=0.6908$ ($6$~dB of squeezing):
\begin{equation}
\ket{r} = \frac{1}{\sqrt{\cosh r}} \sum_{n=0}^{n_c} (-1)^n \frac{\sqrt{(2n)!}}{2^n n!} \tanh^n r \ket{2n}.
\end{equation}
\item Fock state $\rho=\ket{5}\bra{5}$.
\item Three-term Fock superposition state: 
\begin{equation}
\rho = \frac{1}{3}\left(\ket{4}+\ket{5}+\ket{6}\right)\left(\bra{4}+\bra{5}+\bra{6}\right).
\end{equation}
\end{enumerate}

\begin{figure*}[tb!]
\centering
\includegraphics[width=7in]{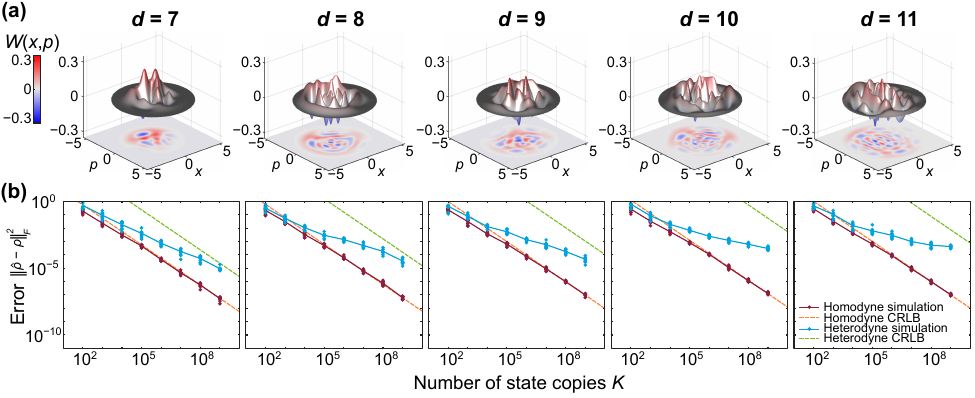}
\caption{Simulation results for random non-Gaussian states with Hilbert space dimension $d\in\{7,8,9,10,11\}$. (a) Ground truth Wigner functions and (b) corresponding estimation errors. Circles denote the Frobenius error for a specific trial, solid lines trace the mean Frobenius error of ten trials, and dashed lines give the CRLB.}
\label{fig:rand2}
\end{figure*}

While all random states tested in Sec.~\ref{sec:random} are both non-Gaussian and highly mixed, States 1--3 here are Gaussian, and States 2--6 are pure, consequently offering distinct features for testing our CFI- and MLE-based analyses. The chosen parameters $\lambda$, $\alpha$, and $r$ ensure that the truncation error, defined as $\epsilon = 1-\sum_{n=0}^{n_c} \braket{n|\rho|n}$, is less than $10^{-3}$, although slight ripples in the  coherent and squeezed Wigner functions plotted in Fig.~\ref{fig:tailored}(a) show that some truncation artifacts remain.

Simulation results for these states appear in Fig.~\ref{fig:tailored}. The ground truth Wigner functions for the first three are positive and Gaussian while the second three all show broad regions of negativity and non-Gaussian features. Interestingly, the CFI for all six states predicts markedly greater tomographic efficiency for homodyne measurements compared to heterodyne, even for the Gaussian examples---a prediction not inconsistent with work suggesting greater efficiency for heterodyne measurements, for here we seek to infer the \emph{full} ($d^2-1$)-length parameter vector $\bt$ rather than the covariance matrix alone~\cite{Rehacek2015, Muller2016, Teo2017}.

Yet with the exception of the homodyne results for the thermal and coherent examples, quantitative agreement between the empirical Frobenius error and the corresponding CRLB is extremely poor. In particular, heterodyne tomography attains vastly lower mean-squared errors---by up to five orders of magnitude in the squeezed example---than the CFI prediction. Given the $d=11$ Hilbert space dimension for all these tests, failure to reach the asymptotic regime---similar to the trends in the $d=11$ random example [Fig.~\ref{fig:rand2}(b)]---could be to blame for the wide heterodyne deviations.

On the other hand, the remaining quantitative discrepancies in the homodyne cases in Fig.~\ref{fig:tailored} are quite surprising, in light of the fact that all random homodyne examples track the CRLB closely [Figs.~\ref{fig:rand1}(b) and \ref{fig:rand2}(b)]. Although we have yet to arrive at a definitive explanation, we suspect that high state purity $\Tr \rho^2$ is responsible at least in part. Whereas the purities of the ten  examples in Sec.~\ref{sec:random} are appreciably smaller than unity, indicating the states in question lie far from the edge of the Bloch hypersphere (specifically, $\Tr\rho^2 \in\set{0.92,0.82,0.85,0.73,0.90,0.85,0.83,0.86,0.87,0.83}$), the coherent, squeezed, Fock, and superposition states in this subsection are exactly on the edge ($\Tr\rho^2=1$), meaning that the asymptotic normality assumed in the CRLB does not hold~\cite{Scholten_2018}.

\section{Discussion}\label{sec:Disc}
\begin{figure*}[t!]
\centering
\includegraphics[width=7in]{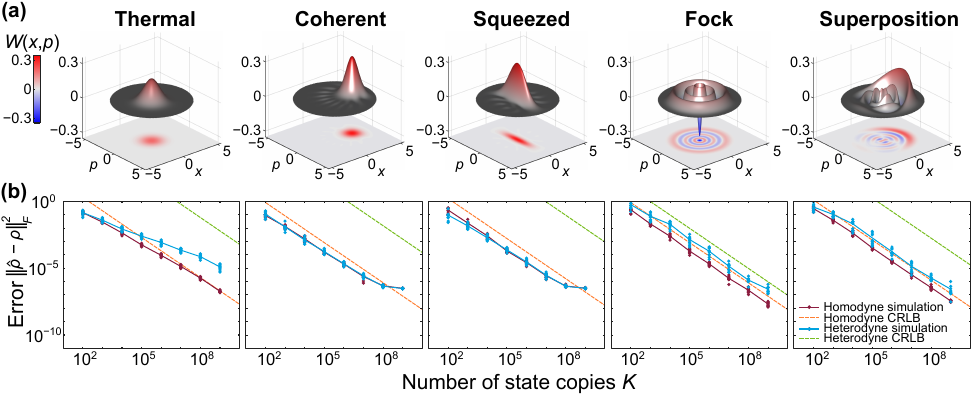}
\caption{Simulation results for common optical states truncated to Hilbert space dimension $d=11$. (a) Ground truth Wigner functions. (b) Estimation errors for ten simulated experiments on each state. See Sec.~\ref{sec:tailored} for state definitions.}
\label{fig:tailored}
\end{figure*}

In general, homodyne tests show good agreement with the CRLB---quantitatively for random states [Figs.~\ref{fig:rand1}(b) and \ref{fig:rand2}(b)], qualitatively for tailored examples [Fig.~\ref{fig:tailored}(b)]. But the mean-squared estimation errors obtained in heterodyne tomography are often orders of magnitude below the CRLB, only appearing to trend toward the CRLB asymptotically. Incidentally, such heterodyne challenges seem strikingly similar to other research threads touching on this measurement modality.
For example, in recent work deriving sample complexity bounds for classical shadow tomography of CV states,
only an exponential scaling bound was found for heterodyne measurements, compared to a polynomial bound in the homodyne case~\cite{Gandhari2023}. Similarly, previous work on model selection in the presence of positivity constraints ($\rho\succeq0$) observed relatively poor agreement in heterodyne tomography under finite measurements, so much so that the authors noted, ``Our numerical experiments with heterodyne tomography show unexpected behavior, indicating that quantum tomography can still surprise, and may violate all asymptotic statistics results'' ~\cite{Scholten_2018}. Our results corroborate and add evidence for what appears to be a general difficulty in matching finite-measurement heterodyne results with asymptotic theory.

Moving forward, it would be valuable to push to even more state copies $K>10^9$ in an effort to approach asymptotic behavior. Currently, the maximum number of measurements is limited primarily by the memory required in the multinomial random number generator, which we found to exhaust all available RAM in our machine for $K=10^{10}$ copies. Significantly more state copies may be possible with more advanced parallelization techniques. 
As another research front, applying our method to computationally useful non-Gaussian states like GKP qubits~\cite{Gottesman2001} could prove insightful, although our results so far strongly suggest homodyne measurements to be preferred. 

Overall, our formalism and simulations provide a valuable 
toolkit for CV tomography, enriching our understanding of  measurements tied to fundamental quasiprobability distributions in quantum mechanics, the Wigner and Husimi $Q$ functions. Although suggesting that homodyne measurements are much more efficient than heterodyne for random non-Gaussian states, our results also reveal that the \emph{degree} of separation is not nearly as significant as suggested by asymptotic theory---%
underlining by example the constant vigilance required when applying asymptotic results to  finite-measurement regimes.

\begin{acknowledgments}
This work was performed in part at Oak Ridge National Laboratory, operated by UT-Battelle for the U.S. Department of Energy under Contract No. DE-AC05-00OR22725. Funding was provided by the National Science Foundation (ECCS-2540189) and the U.S. Department of Energy, Office of Science, Advanced Scientific Computing Research (ERKJ432).
\end{acknowledgments}





\FloatBarrier
\bibliography{BIB}

@article{Pizzimenti2024,
  title = {Optical {Gottesman-Kitaev-Preskill} qubit generation via approximate squeezed coherent state superposition breeding},
  author = {Pizzimenti, Andrew J. and Soh, Daniel},
  journal = {Phys. Rev. A},
  volume = {110},
  issue = {6},
  pages = {062619},
  numpages = {11},
  year = {2024},
  month = {Dec},
  publisher = {American Physical Society},
  doi = {10.1103/PhysRevA.110.062619},
  url = {https://link.aps.org/doi/10.1103/PhysRevA.110.062619}
}

@article{Su2019,
	title = {Conversion of {Gaussian} states to non-{Gaussian} states using photon-number-resolving detectors},
	author = {Su, Daiqin and Myers, Casey R. and Sabapathy, Krishna Kumar},
	journal = {Phys. Rev. A},
	volume = {100},
	issue = {5},
	pages = {052301},
	numpages = {32},
	year = {2019},
	month = {Nov},
	publisher = {American Physical Society},
	doi = {10.1103/PhysRevA.100.052301},
	url = {https://link.aps.org/doi/10.1103/PhysRevA.100.052301}
}

@article{Gagatsos2019,
	title = {Efficient representation of {Gaussian} states for multimode non-{Gaussian} quantum state engineering via subtraction of arbitrary number of photons},
	author = {Gagatsos, Christos N. and Guha, Saikat},
	journal = {Phys. Rev. A},
	volume = {99},
	issue = {5},
	pages = {053816},
	numpages = {10},
	year = {2019},
	month = {May},
	publisher = {American Physical Society},
	doi = {10.1103/PhysRevA.99.053816},
	url = {https://link.aps.org/doi/10.1103/PhysRevA.99.053816}
}

@article{Walschaers2020,
	title = {Practical Framework for Conditional Non-{Gaussian} Quantum State Preparation},
	author = {Walschaers, Mattia and Parigi, Valentina and Treps, Nicolas},
	journal = {PRX Quantum},
	volume = {1},
	issue = {2},
	pages = {020305},
	numpages = {14},
	year = {2020},
	month = {Oct},
	publisher = {American Physical Society},
	doi = {10.1103/PRXQuantum.1.020305},
	url = {https://link.aps.org/doi/10.1103/PRXQuantum.1.020305}
}

@article{Bartlett2002,
	title = {Universal continuous-variable quantum computation: Requirement of optical nonlinearity for photon counting},
	author = {Bartlett, Stephen D. and Sanders, Barry C.},
	journal = {Phys. Rev. A},
	volume = {65},
	issue = {4},
	pages = {042304},
	numpages = {5},
	year = {2002},
	month = {Mar},
	publisher = {American Physical Society},
	doi = {10.1103/PhysRevA.65.042304},
	url = {https://link.aps.org/doi/10.1103/PhysRevA.65.042304}
}

@article{Braunstein2005,
  title = {Quantum information with continuous variables},
  author = {Braunstein, Samuel L. and van Loock, Peter},
  journal = {Rev. Mod. Phys.},
  volume = {77},
  issue = {2},
  pages = {513--577},
  numpages = {0},
  year = {2005},
  month = {Jun},
  publisher = {American Physical Society},
  doi = {10.1103/RevModPhys.77.513},
  url = {https://link.aps.org/doi/10.1103/RevModPhys.77.513}
}

@article{Weedbrook2012,
  title = {Gaussian quantum information},
  author = {Weedbrook, Christian and Pirandola, Stefano and Garc\'{\i}a-Patr\'on, Ra\'ul and Cerf, Nicolas J. and Ralph, Timothy C. and Shapiro, Jeffrey H. and Lloyd, Seth},
  journal = {Rev. Mod. Phys.},
  volume = {84},
  issue = {2},
  pages = {621--669},
  numpages = {0},
  year = {2012},
  month = {May},
  publisher = {American Physical Society},
  doi = {10.1103/RevModPhys.84.621},
  url = {https://link.aps.org/doi/10.1103/RevModPhys.84.621}
}

@article{Lloyd1999,
  title = {Quantum Computation over Continuous Variables},
  author = {Lloyd, Seth and Braunstein, Samuel L.},
  journal = {Phys. Rev. Lett.},
  volume = {82},
  issue = {8},
  pages = {1784--1787},
  numpages = {0},
  year = {1999},
  month = {Feb},
  publisher = {American Physical Society},
  doi = {10.1103/PhysRevLett.82.1784},
  url = {https://link.aps.org/doi/10.1103/PhysRevLett.82.1784}
}

@article{Gottesman2001,
  title = {Encoding a qubit in an oscillator},
  author = {Gottesman, Daniel and Kitaev, Alexei and Preskill, John},
  journal = {Phys. Rev. A},
  volume = {64},
  issue = {1},
  pages = {012310},
  numpages = {21},
  year = {2001},
  month = {Jun},
  publisher = {American Physical Society},
  doi = {10.1103/PhysRevA.64.012310},
  url = {https://link.aps.org/doi/10.1103/PhysRevA.64.012310}
}

@article{Tzitrin2020,
  title = {Progress towards practical qubit computation using approximate {Gottesman--Kitaev--Preskill} codes},
  author = {Tzitrin, Ilan and Bourassa, J. Eli and Menicucci, Nicolas C. and Sabapathy, Krishna Kumar},
  journal = {Phys. Rev. A},
  volume = {101},
  issue = {3},
  pages = {032315},
  numpages = {31},
  year = {2020},
  month = {Mar},
  publisher = {American Physical Society},
  doi = {10.1103/PhysRevA.101.032315},
  url = {https://link.aps.org/doi/10.1103/PhysRevA.101.032315}
}

@article{Knill2001,
  title={A scheme for efficient quantum computation with linear optics},
  author={Knill, Emanuel and Laflamme, Raymond and Milburn, Gerald J},
  journal={Nature},
  volume={409},
  number={6816},
  pages={46--52},
  year={2001},
  publisher={Nature Publishing Group},
  url={https://doi.org/10.1038/35051009}
}

@article{Kok2007,
  title = {Linear optical quantum computing with photonic qubits},
  author = {Kok, Pieter and Munro, W. J. and Nemoto, Kae and Ralph, T. C. and Dowling, Jonathan P. and Milburn, G. J.},
  journal = {Rev. Mod. Phys.},
  volume = {79},
  issue = {1},
  pages = {135--174},
  numpages = {0},
  year = {2007},
  month = {Jan},
  publisher = {American Physical Society},
  doi = {10.1103/RevModPhys.79.135},
  url = {http://link.aps.org/doi/10.1103/RevModPhys.79.135}
}

@article{Hamilton2017,
  title = {{Gaussian} Boson Sampling},
  author = {Hamilton, Craig S. and Kruse, Regina and Sansoni, Linda and Barkhofen, Sonja and Silberhorn, Christine and Jex, Igor},
  journal = {Phys. Rev. Lett.},
  volume = {119},
  issue = {17},
  pages = {170501},
  numpages = {5},
  year = {2017},
  month = {Oct},
  publisher = {American Physical Society},
  doi = {10.1103/PhysRevLett.119.170501},
  url = {https://link.aps.org/doi/10.1103/PhysRevLett.119.170501}
}

@article{Kruse2019,
  title = {Detailed study of {Gaussian} boson sampling},
  author = {Kruse, Regina and Hamilton, Craig S. and Sansoni, Linda and Barkhofen, Sonja and Silberhorn, Christine and Jex, Igor},
  journal = {Phys. Rev. A},
  volume = {100},
  issue = {3},
  pages = {032326},
  numpages = {15},
  year = {2019},
  month = {Sep},
  publisher = {American Physical Society},
  doi = {10.1103/PhysRevA.100.032326},
  url = {https://link.aps.org/doi/10.1103/PhysRevA.100.032326}
}

@article{Zhong2019,
title = {Experimental {Gaussian} Boson sampling},
journal = {Sci. Bull.},
volume = {64},
number = {8},
pages = {511-515},
year = {2019},
issn = {2095-9273},
doi = {https://doi.org/10.1016/j.scib.2019.04.007},
url = {https://www.sciencedirect.com/science/article/pii/S2095927319301938},
author = {Han-Sen Zhong and Li-Chao Peng and Yuan Li and Yi Hu and Wei Li and Jian Qin and Dian Wu and Weijun Zhang and Hao Li and Lu Zhang and Zhen Wang and Lixing You and Xiao Jiang and Li Li and Nai-Le Liu and Jonathan P. Dowling and Chao-Yang Lu and Jian-Wei Pan}
}

@article{Paesani2019,
  title={Generation and sampling of quantum states of light in a silicon chip},
  author={Paesani, Stefano and Ding, Yunhong and Santagati, Raffaele and Chakhmakhchyan, Levon and Vigliar, Caterina and Rottwitt, Karsten and Oxenl{\o}we, Leif K and Wang, Jianwei and Thompson, Mark G and Laing, Anthony},
  journal={Nat. Phys.},
  volume={15},
  number={9},
  pages={925--929},
  year={2019},
  publisher={Nature Publishing Group},
  doi={10.1038/s41567-019-0567-8}
}

@article{Arrazola2021,
  doi = {10.1038/s41586-021-03202-1},
  url = {https://doi.org/10.1038/s41586-021-03202-1},
  year = {2021},
  month = mar,
  publisher = {Springer Science and Business Media {LLC}},
  volume = {591},
  number = {7848},
  pages = {54--60},
  author = {J. M. Arrazola and V. Bergholm and K. Br{\'{a}}dler and T. R. Bromley and M. J. Collins and I. Dhand and A. Fumagalli and T. Gerrits and A. Goussev and L. G. Helt and J. Hundal and T. Isacsson and R. B. Israel and J. Izaac and S. Jahangiri and R. Janik and N. Killoran and S. P. Kumar and J. Lavoie and A. E. Lita and D. H. Mahler and M. Menotti and B. Morrison and S. W. Nam and L. Neuhaus and H. Y. Qi and N. Quesada and A. Repingon and K. K. Sabapathy and M. Schuld and D. Su and J. Swinarton and A. Sz{\'{a}}va and K. Tan and P. Tan and V. D. Vaidya and Z. Vernon and Z. Zabaneh and Y. Zhang},
  title = {Quantum circuits with many photons on a programmable nanophotonic chip},
  journal = {Nature}
}

@article{Sabapathy2019,
  title = {Production of photonic universal quantum gates enhanced by machine learning},
  author = {Sabapathy, Krishna Kumar and Qi, Haoyu and Izaac, Josh and Weedbrook, Christian},
  journal = {Phys. Rev. A},
  volume = {100},
  issue = {1},
  pages = {012326},
  numpages = {10},
  year = {2019},
  month = {Jul},
  publisher = {American Physical Society},
  doi = {10.1103/PhysRevA.100.012326},
  url = {https://link.aps.org/doi/10.1103/PhysRevA.100.012326}
}

@article{Quesada2019,
  title = {Simulating realistic non-{Gaussian} state preparation},
  author = {Quesada, N. and Helt, L. G. and Izaac, J. and Arrazola, J. M. and Shahrokhshahi, R. and Myers, C. R. and Sabapathy, K. K.},
  journal = {Phys. Rev. A},
  volume = {100},
  issue = {2},
  pages = {022341},
  numpages = {10},
  year = {2019},
  month = {Aug},
  publisher = {American Physical Society},
  doi = {10.1103/PhysRevA.100.022341},
  url = {https://link.aps.org/doi/10.1103/PhysRevA.100.022341}
}

@article{Walschaers2021,
  title = {Non-{Gaussian} Quantum States and Where to Find Them},
  author = {Walschaers, Mattia},
  journal = {PRX Quantum},
  volume = {2},
  issue = {3},
  pages = {030204},
  numpages = {68},
  year = {2021},
  month = {Sep},
  publisher = {American Physical Society},
  doi = {10.1103/PRXQuantum.2.030204},
  url = {https://link.aps.org/doi/10.1103/PRXQuantum.2.030204}
}

@PREAMBLE{
 "\providecommand{\noopsort}[1]{}" 
 # "\providecommand{\singleletter}[1]{#1}%" 
}

@article{Pizzimenti2021,
  title={Non-{Gaussian} photonic state engineering with the quantum frequency processor},
  author={Pizzimenti, Andrew J and Lukens, Joseph M and Lu, Hsuan-Hao and Peters, Nicholas A and Guha, Saikat and Gagatsos, Christos N},
  journal={Phys. Rev. A},
  volume={104},
  number={6},
  pages={062437},
  year={2021},
  publisher={APS},
  url={https://doi.org/10.1103/PhysRevA.104.062437}
}

@article{Barbosa2003,
  title = {Secure Communication Using Mesoscopic Coherent States},
  author = {Barbosa, Geraldo A. and Corndorf, Eric and Kumar, Prem and Yuen, Horace P.},
  journal = {Phys. Rev. Lett.},
  volume = {90},
  issue = {22},
  pages = {227901},
  numpages = {4},
  year = {2003},
  month = {Jun},
  publisher = {American Physical Society},
  doi = {10.1103/PhysRevLett.90.227901},
  url = {https://link.aps.org/doi/10.1103/PhysRevLett.90.227901}
}

@article{Grosshans2002,
  title = {Continuous Variable Quantum Cryptography Using Coherent States},
  author = {Grosshans, Fr\'ed\'eric and Grangier, Philippe},
  journal = {Phys. Rev. Lett.},
  volume = {88},
  issue = {5},
  pages = {057902},
  numpages = {4},
  year = {2002},
  month = {Jan},
  publisher = {American Physical Society},
  doi = {10.1103/PhysRevLett.88.057902},
  url = {https://link.aps.org/doi/10.1103/PhysRevLett.88.057902}
}

@article{Laudenbach2018,
author = {Laudenbach, Fabian and Pacher, Christoph and Fung, Chi-Hang Fred and Poppe, Andreas and Peev, Momtchil and Schrenk, Bernhard and Hentschel, Michael and Walther, Philip and Hübel, Hannes},
title = {Continuous-Variable Quantum Key Distribution with {Gaussian} Modulation---The Theory of Practical Implementations},
journal = {Adv. Quantum Technol.},
volume = {1},
number = {1},
pages = {1800011},
doi = {https://doi.org/10.1002/qute.201800011},
url = {https://onlinelibrary.wiley.com/doi/abs/10.1002/qute.201800011},
year = {2018}
}

@article{Lawrie2019,
  title = {Quantum Sensing with Squeezed Light},
  volume = {6},
  ISSN = {2330-4022},
  url = {http://dx.doi.org/10.1021/acsphotonics.9b00250},
  DOI = {10.1021/acsphotonics.9b00250},
  number = {6},
  journal = {ACS Photon.},
  publisher = {American Chemical Society (ACS)},
  author = {Lawrie,  B. J. and Lett,  P. D. and Marino,  A. M. and Pooser,  R. C.},
  year = {2019},
  month = may,
  pages = {1307--1318}
}

@article{Aasi2013,
  title = {Enhanced sensitivity of the {LIGO} gravitational wave detector by using squeezed states of light},
  volume = {7},
  ISSN = {1749-4893},
  url = {http://dx.doi.org/10.1038/nphoton.2013.177},
  DOI = {10.1038/nphoton.2013.177},
  number = {8},
  journal = {Nat. Photon.},
  publisher = {Springer Science and Business Media LLC},
  author = {{The LIGO Scientific Collaboration}},
  year = {2013},
  month = jul,
  pages = {613--619}
}

@article{Yu2020,
  title = {Quantum correlations between light and the kilogram-mass mirrors of {LIGO}},
  volume = {583},
  ISSN = {1476-4687},
  url = {http://dx.doi.org/10.1038/s41586-020-2420-8},
  DOI = {10.1038/s41586-020-2420-8},
  number = {7814},
  journal = {Nature},
  publisher = {Springer Science and Business Media LLC},
  author = {{The LIGO Scientific Collaboration}},
  year = {2020},
  month = jul,
  pages = {43--47}
}

@article{Zhong2020,
author = {Han-Sen Zhong  and Hui Wang  and Yu-Hao Deng  and Ming-Cheng Chen  and Li-Chao Peng  and Yi-Han Luo  and Jian Qin  and Dian Wu  and Xing Ding  and Yi Hu  and Peng Hu  and Xiao-Yan Yang  and Wei-Jun Zhang  and Hao Li  and Yuxuan Li  and Xiao Jiang  and Lin Gan  and Guangwen Yang  and Lixing You  and Zhen Wang  and Li Li  and Nai-Le Liu  and Chao-Yang Lu  and Jian-Wei Pan },
title = {Quantum computational advantage using photons},
journal = {Science},
volume = {370},
number = {6523},
pages = {1460-1463},
year = {2020},
doi = {10.1126/science.abe8770},
URL = {https://www.science.org/doi/abs/10.1126/science.abe8770}}

@article{Zhong2021,
  title = {Phase-Programmable {Gaussian} Boson Sampling Using Stimulated Squeezed Light},
  author = {Zhong, Han-Sen and Deng, Yu-Hao and Qin, Jian and Wang, Hui and Chen, Ming-Cheng and Peng, Li-Chao and Luo, Yi-Han and Wu, Dian and Gong, Si-Qiu and Su, Hao and Hu, Yi and Hu, Peng and Yang, Xiao-Yan and Zhang, Wei-Jun and Li, Hao and Li, Yuxuan and Jiang, Xiao and Gan, Lin and Yang, Guangwen and You, Lixing and Wang, Zhen and Li, Li and Liu, Nai-Le and Renema, Jelmer J. and Lu, Chao-Yang and Pan, Jian-Wei},
  journal = {Phys. Rev. Lett.},
  volume = {127},
  issue = {18},
  pages = {180502},
  numpages = {9},
  year = {2021},
  month = {Oct},
  publisher = {American Physical Society},
  doi = {10.1103/PhysRevLett.127.180502},
  url = {https://link.aps.org/doi/10.1103/PhysRevLett.127.180502}
}

@article{Thekkadath2022,
  title = {Experimental Demonstration of {Gaussian} Boson Sampling with Displacement},
  author = {Thekkadath, G.S. and Sempere-Llagostera, S. and Bell, B.A. and Patel, R.B. and Kim, M.S. and Walmsley, I.A.},
  journal = {PRX Quantum},
  volume = {3},
  issue = {2},
  pages = {020336},
  numpages = {12},
  year = {2022},
  month = {May},
  publisher = {American Physical Society},
  doi = {10.1103/PRXQuantum.3.020336},
  url = {https://link.aps.org/doi/10.1103/PRXQuantum.3.020336}
}

@article{Lvovsky2001,
  title = {Quantum State Reconstruction of the Single-Photon {Fock} State},
  author = {Lvovsky, A. I. and Hansen, H. and Aichele, T. and Benson, O. and Mlynek, J. and Schiller, S.},
  journal = {Phys. Rev. Lett.},
  volume = {87},
  issue = {5},
  pages = {050402},
  numpages = {4},
  year = {2001},
  month = {Jul},
  publisher = {American Physical Society},
  doi = {10.1103/PhysRevLett.87.050402},
  url = {https://link.aps.org/doi/10.1103/PhysRevLett.87.050402}
}

@article{Lvovsky2009,
  title = {Continuous-variable optical quantum-state tomography},
  author = {Lvovsky, A. I. and Raymer, M. G.},
  journal = {Rev. Mod. Phys.},
  volume = {81},
  issue = {1},
  pages = {299--332},
  numpages = {0},
  year = {2009},
  month = {Mar},
  publisher = {American Physical Society},
  doi = {10.1103/RevModPhys.81.299},
  url = {https://link.aps.org/doi/10.1103/RevModPhys.81.299}
}

@article{Teo2017,
  title = {Superiority of heterodyning over homodyning: An assessment with quadrature moments},
  author = {Teo, Y. S. and M\"uller, C. R. and Jeong, H. and Hradil, Z. and \ifmmode \check{R}\else \v{R}\fi{}eh\'a\ifmmode \check{c}\else \v{c}\fi{}ek, J. and S\'anchez-Soto, L. L.},
  journal = {Phys. Rev. A},
  volume = {95},
  issue = {4},
  pages = {042322},
  numpages = {17},
  year = {2017},
  month = {Apr},
  publisher = {American Physical Society},
  doi = {10.1103/PhysRevA.95.042322},
  url = {https://link.aps.org/doi/10.1103/PhysRevA.95.042322}
}

@article{Muller2016,
  title = {Evading Vacuum Noise: Wigner Projections or {Husimi} Samples?},
  author = {M\"{u}ller, C. R. and Peuntinger, C. and Dirmeier, T. and Khan, I. and Vogl, U. and Marquardt, Ch. and Leuchs, G. and S\'anchez-Soto, L. L. and Teo, Y. S. and Hradil, Z. and \ifmmode \check{R}\else \v{R}\fi{}eh\'a\ifmmode \check{c}\else \v{c}\fi{}ek, J.},
  journal = {Phys. Rev. Lett.},
  volume = {117},
  issue = {7},
  pages = {070801},
  numpages = {6},
  year = {2016},
  month = {Aug},
  publisher = {American Physical Society},
  doi = {10.1103/PhysRevLett.117.070801},
  url = {https://link.aps.org/doi/10.1103/PhysRevLett.117.070801}
}

@article{Rehacek2015,
  title = {Surmounting intrinsic quantum-measurement uncertainties in {Gaussian}-state tomography with quadrature squeezing},
  volume = {5},
  ISSN = {2045-2322},
  url = {http://dx.doi.org/10.1038/srep12289},
  DOI = {10.1038/srep12289},
  number = {1},
  pages={12289},
  journal = {Sci. Rep.},
  publisher = {Springer Science and Business Media LLC},
  author = {Řeháček,  Jaroslav and Teo,  Yong Siah and Hradil,  Zdeněk and Wallentowitz,  Sascha},
  year = {2015},
  month = jul 
}

@article{Richter1998,
author = { Th.   Richter },
title = {Determination of photon statistics and density matrix from double homodyne detection measurements},
journal = {J. Mod. Opt.},
volume = {45},
number = {8},
pages = {1735-1749},
year  = {1998},
publisher = {Taylor & Francis},
doi = {10.1080/09500349808230666}
}

@article{Ourjoumtsev2006,
author = {Alexei Ourjoumtsev  and Rosa Tualle-Brouri  and Julien Laurat  and Philippe Grangier },
title = {Generating Optical {Schrödinger} Kittens for Quantum Information Processing},
journal = {Science},
volume = {312},
number = {5770},
pages = {83--86},
year = {2006},
doi = {10.1126/science.1122858},
URL = {https://www.science.org/doi/abs/10.1126/science.1122858}}

@article{Neergaard2006,
  title = {Generation of a Superposition of Odd Photon Number States for Quantum Information Networks},
  author = {Neergaard-Nielsen, J. S. and Nielsen, B. Melholt and Hettich, C. and M\o{}lmer, K. and Polzik, E. S.},
  journal = {Phys. Rev. Lett.},
  volume = {97},
  issue = {8},
  pages = {083604},
  numpages = {4},
  year = {2006},
  month = {Aug},
  publisher = {American Physical Society},
  doi = {10.1103/PhysRevLett.97.083604},
  url = {https://link.aps.org/doi/10.1103/PhysRevLett.97.083604}
}

@article{Wakui2007,
author = {Kentaro Wakui and Hiroki Takahashi and Akira Furusawa and Masahide Sasaki},
journal = {Opt. Express},
number = {6},
pages = {3568--3574},
publisher = {Optica Publishing Group},
title = {Photon subtracted squeezed states generated with periodically poled {KTiOPO$_4$}},
volume = {15},
month = {Mar},
year = {2007},
url = {https://opg.optica.org/oe/abstract.cfm?URI=oe-15-6-3568},
doi = {10.1364/OE.15.003568},
}

@article{Ourjoumtsev2007,
  title = {Generation of optical ‘{Schr\"{o}dinger cats}’ from photon number states},
  volume = {448},
  ISSN = {1476-4687},
  url = {http://dx.doi.org/10.1038/nature06054},
  DOI = {10.1038/nature06054},
  number = {7155},
  journal = {Nature},
  publisher = {Springer Science and Business Media LLC},
  author = {Ourjoumtsev,  Alexei and Jeong,  Hyunseok and Tualle-Brouri,  Rosa and Grangier,  Philippe},
  year = {2007},
  month = aug,
  pages = {784--786}
}

@article{Takahashi2008,
  title = {Generation of Large-Amplitude Coherent-State Superposition via Ancilla-Assisted Photon Subtraction},
  author = {Takahashi, Hiroki and Wakui, Kentaro and Suzuki, Shigenari and Takeoka, Masahiro and Hayasaka, Kazuhiro and Furusawa, Akira and Sasaki, Masahide},
  journal = {Phys. Rev. Lett.},
  volume = {101},
  issue = {23},
  pages = {233605},
  numpages = {4},
  year = {2008},
  month = {Dec},
  publisher = {American Physical Society},
  doi = {10.1103/PhysRevLett.101.233605},
  url = {https://link.aps.org/doi/10.1103/PhysRevLett.101.233605}
}

@article{Gerrits2010,
  title = {Generation of optical coherent-state superpositions by number-resolved photon subtraction from the squeezed vacuum},
  author = {Gerrits, Thomas and Glancy, Scott and Clement, Tracy S. and Calkins, Brice and Lita, Adriana E. and Miller, Aaron J. and Migdall, Alan L. and Nam, Sae Woo and Mirin, Richard P. and Knill, Emanuel},
  journal = {Phys. Rev. A},
  volume = {82},
  issue = {3},
  pages = {031802},
  numpages = {4},
  year = {2010},
  month = {Sep},
  publisher = {American Physical Society},
  doi = {10.1103/PhysRevA.82.031802},
  url = {https://link.aps.org/doi/10.1103/PhysRevA.82.031802}
}

@article{Ra2020,
  title = {Non-{Gaussian} quantum states of a multimode light field},
  volume = {16},
  ISSN = {1745-2481},
  url = {http://dx.doi.org/10.1038/s41567-019-0726-y},
  DOI = {10.1038/s41567-019-0726-y},
  number = {2},
  journal = {Nat. Phys.},
  publisher = {Springer Science and Business Media LLC},
  author = {Ra, Young-Sik and Dufour, Adrien and Walschaers, Mattia and Jacquard,  Clément and Michel, Thibault and Fabre,  Claude and Treps,  Nicolas},
  year = {2020},
  pages = {144--147}
}

@article{KIMURA2003339,
title = {The {Bloch} vector for {$N$}-level systems},
journal = {Phys. Lett. A},
volume = {314},
number = {5},
pages = {339-349},
year = {2003},
issn = {0375-9601},
doi = {https://doi.org/10.1016/S0375-9601(03)00941-1},
url = {https://www.sciencedirect.com/science/article/pii/S0375960103009411},
author = {Gen Kimura}
}

@article{RevModPhys.81.299,
  title = {Continuous-variable optical quantum-state tomography},
  author = {Lvovsky, A. I. and Raymer, M. G.},
  journal = {Rev. Mod. Phys.},
  volume = {81},
  issue = {1},
  pages = {299--332},
  numpages = {0},
  year = {2009},
  month = {Mar},
  publisher = {American Physical Society},
  doi = {10.1103/RevModPhys.81.299},
  url = {https://link.aps.org/doi/10.1103/RevModPhys.81.299}
}

@article{PhysRevA.64.052312,
  title = {Measurement of qubits},
  author = {James, Daniel F. V. and Kwiat, Paul G. and Munro, William J. and White, Andrew G.},
  journal = {Phys. Rev. A},
  volume = {64},
  issue = {5},
  pages = {052312},
  numpages = {15},
  year = {2001},
  month = {Oct},
  publisher = {American Physical Society},
  doi = {10.1103/PhysRevA.64.052312},
  url = {https://link.aps.org/doi/10.1103/PhysRevA.64.052312}
}

@book{scully1997quantum,
  title={Quantum Optics},
  author={Scully, Marlan O and Zubairy, M Suhail},
  year={1997},
  publisher={Cambridge},
  doi={10.1017/CBO9780511813993}
}

@article{PhysRevA.40.2847,
  title = {Determination of quasiprobability distributions in terms of probability distributions for the rotated quadrature phase},
  author = {Vogel, K. and Risken, H.},
  journal = {Phys. Rev. A},
  volume = {40},
  issue = {5},
  pages = {2847--2849},
  numpages = {0},
  year = {1989},
  month = {Sep},
  publisher = {American Physical Society},
  doi = {10.1103/PhysRevA.40.2847},
  url = {https://link.aps.org/doi/10.1103/PhysRevA.40.2847}
}

@article{Walker1986MultiportHD,
  title={Multiport homodyne detection near the quantum noise limit},
  author={N. G. Walker and John E. Carroll},
  journal={Opt. Quantum Electron.},
  year={1986},
  volume={18},
  pages={355-363},
  url={https://api.semanticscholar.org/CorpusID:120412083}
}

@article{Chapman:22,
author = {Joseph C. Chapman and Joseph M. Lukens and Bing Qi and Raphael C. Pooser and Nicholas A. Peters},
journal = {Opt. Express},
number = {9},
pages = {15184--15200},
publisher = {Optica Publishing Group},
title = {Bayesian homodyne and heterodyne tomography},
volume = {30},
year = {2022},
url = {https://opg.optica.org/oe/abstract.cfm?URI=oe-30-9-15184},
doi = {10.1364/OE.456597},
}

@book{walls2008quantum,
  title={Quantum Optics},
  author={Walls, D. F. and Milburn, G. J.},
  isbn={9783540285731},
  lccn={2007936291},
  edition={3rd},
  doi={10.1007/978-3-031-84177-4},
  year={2025},
  publisher={Springer}
}

@article{Scholten_2018,
doi = {10.1088/1367-2630/aaa7e2},
url = {https://dx.doi.org/10.1088/1367-2630/aaa7e2},
year = {2018},
month = {feb},
publisher = {IOP Publishing},
volume = {20},
number = {2},
pages = {023050},
author = {Travis L Scholten and Robin Blume-Kohout},
title = {Behavior of the maximum likelihood in quantum state tomography},
journal = {New J. Phys.}
}

@article{Bertlmann2008,
doi = {10.1088/1751-8113/41/23/235303},
url = {https://dx.doi.org/10.1088/1751-8113/41/23/235303},
year = {2008},
month = {may},
publisher = {},
volume = {41},
number = {23},
pages = {235303},
author = {Reinhold A Bertlmann and Philipp Krammer},
title = {Bloch vectors for qudits},
journal = {J. Phys. A: Math. Theor.}
}

@article{Acharya2019,
doi = {10.1088/1751-8121/ab1958},
url = {https://dx.doi.org/10.1088/1751-8121/ab1958},
year = {2019},
month = {may},
publisher = {IOP Publishing},
volume = {52},
number = {23},
pages = {234001},
author = {Acharya, Anirudh and Kypraios, Theodore and Guţă, Mădălin},
title = {A comparative study of estimation methods in quantum tomography},
journal = {J. Phys. A: Math. Theor.},
}

@article{Acharya2016,
doi = {10.1088/1367-2630/18/4/043018},
url = {https://dx.doi.org/10.1088/1367-2630/18/4/043018},
year = {2016},
month = {apr},
publisher = {IOP Publishing},
volume = {18},
number = {4},
pages = {043018},
author = {Acharya, Anirudh and Kypraios, Theodore and Guţă, Mădălin},
title = {Statistically efficient tomography of low rank states with incomplete measurements},
journal = {New J. Phys.}
}

@article{Lvovsky2004,
	doi = {10.1088/1464-4266/6/6/014},
	url = {https://doi.org/10.1088/1464-4266/6/6/014},
	year = 2004,
	publisher = {{IOP} Publishing},
	volume = {6},
	number = {6},
	pages = {S556--S559},
	author = {A I Lvovsky},
	title = {Iterative maximum-likelihood reconstruction in quantum homodyne tomography},
	journal = {J. Opt. B: Quantum Semiclass. Opt.}
}

@article{Larsen2025,
  title = {Integrated photonic source of {Gottesman--Kitaev--Preskill} qubits},
  ISSN = {1476-4687},
  url = {http://dx.doi.org/10.1038/s41586-025-09044-5},
  volume = {642},
  pages = {587--591},
  DOI = {10.1038/s41586-025-09044-5},
  journal = {Nature},
  publisher = {Springer Science and Business Media LLC},
  author = {Larsen,  M. V. and Bourassa,  J. E. and Kocsis,  S. and Tasker,  J. F. and Chadwick,  R. S. and González-Arciniegas,  C. and Hastrup,  J. and Lopetegui-González,  C. E. and Miatto,  F. M. and Motamedi,  A. and Noro,  R. and Roeland,  G. and Baby,  R. and Chen,  H. and Contu,  P. and Di Luch,  I. and Drago,  C. and Giesbrecht,  M. and Grainge,  T. and Krasnokutska,  I. and Menotti,  M. and Morrison,  B. and Puviraj,  C. and Rezaei Shad,  K. and Hussain,  B. and McMahon,  J. and Ortmann,  J. E. and Collins,  M. J. and Ma,  C. and Phillips,  D. S. and Seymour,  M. and Tang,  Q. Y. and Yang,  B. and Vernon,  Z. and Alexander,  R. N. and Mahler,  D. H.},
  year = {2025},
  month = jun 
}

@article{Leonhardt1993,
  title = {Phase measurement and {$Q$} function},
  author = {Leonhardt, U. and Paul, H.},
  journal = {Phys. Rev. A},
  volume = {47},
  issue = {4},
  pages = {R2460--R2463},
  numpages = {0},
  year = {1993},
  month = {Apr},
  publisher = {American Physical Society},
  doi = {10.1103/PhysRevA.47.R2460},
  url = {https://link.aps.org/doi/10.1103/PhysRevA.47.R2460}
}

@article{Gandhari2023,
      title={Precision Bounds on Continuous-Variable State Tomography using Classical Shadows}, 
      author={Srilekha Gandhari and Victor V. Albert and Thomas Gerrits and Jacob M. Taylor and Michael J. Gullans},
      year={2023},
      journal={arXiv:2211.05149},
      url={https://doi.org/10.48550/arXiv.2211.05149},
}

@BOOK{Orszag2024,
  title     = "Quantum Optics: Including Noise Reduction, Trapped Ions, Quantum Trajectories, and Decoherence",
  author    = "Orszag, Miguel",
  publisher = "Springer",
  edition   =  "4",
  month     =  "jun",
  year      =  "2024",
  doi = "10.1007/978-3-031-54853-6"
}

@article{Guta2011,
  title = {Fisher information and asymptotic normality in system identification for quantum {Markov} chains},
  author = {M\u{a}d\u{a}lin Gu\c{t}\u{a}},
  journal = {Phys. Rev. A},
  volume = {83},
  issue = {6},
  pages = {062324},
  numpages = {9},
  year = {2011},
  publisher = {American Physical Society},
  doi = {10.1103/PhysRevA.83.062324}
}
\end{document}